\documentclass[final,5p,fleqn]{elsarticle}
\usepackage{amsmath,amssymb}
\usepackage{xcolor}
\usepackage{graphicx}
\newcommand{\lipt}{\mbox{$^{11}{\rm Li}(p,t)^{9}{\rm Li}$}}
\newcommand{\hept}{\mbox{$^{6}{\rm He}(p,t)^{4}{\rm He}$}}
\newcommand{\hep}{\mbox{$^{6}$He+p}}
\newcommand{\lit}{\mbox{$^{9}$Li+t}}
\newcommand{\lip}{\mbox{$^{11}$Li+p}}

\newcommand{\elab}{\mbox{$E_{\rm lab}$}}
\newcommand{\ecm}{\mbox{$E_{\rm c.m.}$}}

\journal{Physics Letters B}

\begin{document}
\begin{frontmatter}
	\title{Microscopic study of halo nuclei through $(p,t)$ reactions}
	\author[PNTPM]{P. Descouvemont}
	\ead{pierre.descouvemont@ulb.be}
\address[PNTPM]{D\'epartement de Physique, C.P. 229,
	Universit\'e Libre de Bruxelles (ULB), B 1050 Brussels, Belgium}

\begin{abstract}
We analyze $(p,t)$ two-neutron transfer reactions in a semi-microscopic model. The overlap integrals of the target nucleus are calculated in a microscopic cluster model. The Resonating Group Method (RGM) assumes a cluster structure of the nucleus, and is well adapted to halo nuclei since the long-range part of the wave function is accurately described. We focus on $(p,t)$ reactions involving $^6$He and $^{11}$Li, which are well known core+n+n halo nuclei. The RGM is based on a nucleon-nucleon interaction, and therefore does not involve any fitting procedure. It also provides overlap integrals of excited states of the core nucleus. We present overlap integrals and spectroscopic factors of $^6$He and $^{11}$Li. We compute the $\hept$ and $\lipt$ cross sections at the DWBA, and compare them with experiments. For $^{11}$Li we also determine the  $\lipt^*$ cross section which involves the first excited states of $^{9}$Li. A fair agreement with experiment is obtained, considering that no parameter is adjusted. 
\end{abstract}

\end{frontmatter}

\section{Introduction}
Halo nuclei represent an active field in nuclear physics. Given the low neutron separation energy, these exotic nuclei are close to the driplines, and present unusual properties, such as a large radius \cite{THH85} or soft E1 excitations \cite{Na23}. The development of  radioactive beams has permitted the investigation of halo nuclei through various reaction mechanisms \cite{TSK13}. Elastic scattering, breakup, and fusion are among the most used techniques \cite{CGD15}. These experiments,
however, do not provide a direct access to the nuclear spectroscopy. Scattering models, linking the description of the colliding nuclei and the reaction mechanism, are therefore necessary.

The interest of $(p,t)$ reactions as a spectroscopic tool is well known \cite{Gl65,IKY91}. This two-neutron transfer process has been shown to provide valuable information on neutron-rich nuclei (see Ref.\ \cite{PIB13} and references therein). In $(p,t)$ reactions, the target nucleus is seen as a core surrounded by the two transferred neutrons, and
therefore provides an efficient tool to probe the structure of halo nuclei. Two-neutron transfer data have been obtained on light exotic nuclei, such as $^6$He \cite{WFR99, GRD05}, $^8$He \cite{MLM12}, or
$^{11}$Li \cite{TAB08}.

Calculations of $(p,t)$ cross sections on stable nuclei started long ago (see, for example, Refs.\ \cite{Gl63,Ba70}). 
A detailed description of the literature is presented in Ref.\ \cite{PIB13}. A three-body model of the halo nucleus was used in Ref.\ \cite{OZV99} for $\hept$ and in Ref.\ \cite{De21} for $\lipt$. The influence of the transfer reactions on elastic scattering was investigated in Ref.\ \cite{RGR07}. In Ref.\ \cite{Ti01}, the author discusses the normalization of the $\hept$ cross section with shell-model overlap functions.

An aim in the present work is to study $(p,t)$ reactions in a semi-microscopic model, where overlap integrals are obtained in the Resonating Group Method (RGM, see Ref.\ \cite{Ho77}). In this approach, the Pauli principle is exactly taken into account. Even if a neutron cannot be considered as a genuine cluster, a cluster model is well adapted to describe halo nuclei since the long-range part of the wave function can be accurately described. Another advantage
is that core excited states can be included. The method and its application to several light halo nuclei are presented in Ref.\ \cite{De23}. In $(p,t)$ reactions, the scattering wave functions on the entrance and exit channels are obtained from an optical potential.

The text is organized as follows. First, we briefly describe the RGM, and emphasize the calculation of the overlap integrals. Then we show how to determine the transfer scattering matrices by using the DWBA. Applications to the $\hept$ and $\lipt$ reactions are presented.

\section{Overlap integrals in the RGM}
Our main aim is to investigate $(p,t)$ reactions with a microscopic description of the target. To this end, the Hamiltonian of the target nucleus, involving $A$ nucleons, is written as
\begin{eqnarray}
	H_A = \sum_{i=1}^A t_i + \sum_{i<j}^A v_{ij},
\label{eq1}
\end{eqnarray}
where $t_i$ is the kinetic energy of nucleon $i$, and $v_{ij}$ an effective nucleon-nucleon interaction. It involves a central nuclear term, a spin-orbit component, and the Coulomb interaction. For exotic nuclei, such as $^{11}$Li, the Schr\"{o}dinger equation associated with Hamiltonian \eqref{eq1} does not have an exact solution. We use the Resonating Group Method (RGM) \cite{TLT78,Ho77} which assumes that the nucleons are grouped in clusters. This technique is well adapted to halo nuclei, which present a marked core + neutron structure. 

In a three-cluster approach, the $A$-nucleon wave function is therefore (schematically) written as
\begin{eqnarray}
	\Psi_A = \sum_c \mathcal{A}  \Phi^c_{A-2} \Phi_n \Phi_n G_c(\pmb{r_x}, \pmb{r_y}) ,
\label{eq2}
\end{eqnarray}
where $\mathcal{A}$ is the antisymmetrization operator, and $\pmb{r_x}$ and $\pmb{r_y}$ are, respectively, the neutron-neutron distance, and the relative coordinate between the core and the two-neutron system. For the sake of clarity we omit the spin and parity, as well as the various angular-momentum couplings. In Eq.\ \eqref{eq2}, $\Phi^c_{A-2}$ and $\Phi_n$ are the core wave function and a neutron spinor, and $G_c(\pmb{r_x}, \pmb{r_y})$ is a relative function. The summation runs over different core states, labeled by $c$.  Typical examples of halo nuclei are the $^6$He and $^{11}$Li nuclei, which involve an $\alpha$ and $^9$Li core, respectively. 

The core wave functions $\Phi^c_{A-2}$ are defined in the shell model as
\begin{eqnarray}
	\Phi^c_{A-2} = \sum_i d_i^c \phi_i,
\label{eq3}
\end{eqnarray}
where $\phi_i$ are Slater determinants involving $A-2$ harmonic oscillator orbitals. Coefficients $ d_i^c$ are obtained from the diagonalization of the angular-momentum and isospin operators \cite{DD12}. The core wave functions \eqref{eq3} are therefore characterized by spin, parity and isospin quantum numbers, as expected from the physics of the problem. These indices, however, are omitted for the sake of clarity in this general presentation. 

The shell-model description of the $\alpha$ particle is very simple (4 nucleons in the 0s shell) and a microscopic cluster description of $^6$He has been used by several authors (see, for example, Refs.\ \cite{FKK96,ASL99,BMB02}). 
In contrast, a microscopic approach of the $^{11}$Li nucleus is more difficult \cite{VSL02,De19}. A $p$-shell description of the $^9$Li core involves 90 Slater determinants $\phi_i$. The main difficulty arises from angular momentum projection of the total wave function \eqref{eq2}. This projection involves 7-dimension integrals, which must be performed numerically. The calculation is fast for $^6$He, but highly time consuming for heavier systems such as $^{11}$Li.

In practice, we express the target wave functions \eqref{eq2} by using the Generator Coordinate Method, where it can be written as a superposition of projected Slater determinants (see, for example, Refs.\ \cite{WT77,Ho77,DD12} for details). For three-cluster systems, we adopt the hyperspherical coordinates, where the six coordinates $\pmb{r_x}$ and $\pmb{r_y}$ are replaced by the hyperradius, the hyperangle, and the 4 angles associated with $\pmb{r_x}$ and $\pmb{r_y}$. In this approach, the wave function \eqref{eq2} is expanded over hyperspherical harmonics involving the hypermoment $K$ \cite{Li95,GKV11}.

A significant improvement of the RGM, compared to non-microscopic models, is that excited states of the core are naturally included. It was shown in Refs.\ \cite{De19,De23} that core excitations play an important role in the description of $^{11}$Li. The $^9$Li core presents a 1/2$^-$ excited state at low energy (2.69 MeV) and is expected to affect the $^{11}$Li wave function, and therefore the $\lipt$ cross section.

An important quantity in transfer cross sections is the overlap integral \cite{Ti14}. In a three-body model, the two-neutron overlap integral in channel $c$ is defined as
\begin{eqnarray}
	I_c(\pmb{r_x}, \pmb{r_y}) = \langle  \Phi^c_{A-2}\Phi_n \Phi_n \vert \Psi_A \rangle,
\label{eq4}
\end{eqnarray}
and is different from the relative function $G_c(\pmb{r_x}, \pmb{r_y})$ owing to the presence of the antisymmetrizer in the total wave function \eqref{eq2}. The numerical calculation of the overlap integrals is described in Ref.\ \cite{De23}, where we follow a method developed by Varga and Lovas \cite{VL88}. In practice, $I_c(\pmb{r_x}, \pmb{r_y})$ is expanded in hyperspherical functions and is used as an input in the $(p,t)$ cross sections. The calculation of transfer cross sections to excited states of the residual (core) nucleus is possible with the overlap integrals associated with various channels $c$. This effect is important in the $\lipt$ reaction, where the cross section to the $^9$Li excited state has been measured \cite{TAB08}.

From the overlap integral \eqref{eq4}, we define the radial amplitude as
\begin{eqnarray}
P_c(r_x, r_y)=\int \vert 	I_c(\pmb{r_x}, \pmb{r_y})\vert ^2 d\Omega_x d\Omega_y,
\label{eq5}
\end{eqnarray}
which provides the spectroscopic factor in channel $c$ as
\begin{eqnarray}
	\mathcal{S}_{c} = \int P_c(r_x, r_y)dr_x\, dr_y.
\label{eq6}
\end{eqnarray}
The definition of the spectroscopic factor is well known for two-body systems, but can be extended to three (or more) body systems. This quantity naturally arises in transfer reactions \cite{Au70b,Sa83}.

\section{DWBA calculation of $(p,t)$ cross sections}
We now consider a $(p,t)$ reaction on a $A$-nucleon system composed of a core and of two surrounding neutrons. The initial scattering wave function at energy $E_i$ and in partial wave $J\pi$ reads
\begin{align}
\Psi_i^{JM\pi}(E_i,\pmb{R})=&\left[ [\Psi_A \otimes \Phi_p]^{S_i} \otimes Y_{L_i}(\Omega_R)\right] ^{JM} \nonumber \\
&\times \chi^{J\pi}_{S_i L_i}(E_i,R),
\label{eq7}
\end{align}
where $\pmb{R}$ is the relative coordinate and $L_i$ the orbital angular momentum. In our approach, the target nucleus is described in the RGM with definition \eqref{eq2}. 

Similarly, in the exit channel, the scattering wave function involving the state $c$ of the core nucleus, is defined by
\begin{align}
\Psi_{f,c}^{JM\pi}(E_f,\pmb{R'})=&\left[ [\Phi^c_{A-2} \otimes \Phi_t]^{S_f} \otimes Y_{L_f}(\Omega_{R'})\right] ^{JM} \nonumber \\
&\times \chi^{J\pi}_{S_f L_f,c}(E_f,R'),
\label{eq8}
\end{align}
where $\pmb{R'}$ is the relative coordinate between the triton and the residual nucleus, and $E_f$ the relative energy in the exit channel. Notice that it can be in the ground state, but also in excited states. This property will allow us to study the $\lipt^*$ reaction to the $^9$Li first excited state. In Eqs.\ (\ref{eq7},\ref{eq8}), the relative functions are obtained from optical potentials. The channel spins are denoted as $S_i$ and $S_f$.

From the scattering wave functions, we compute the transfer scattering matrices at the DWBA from
\begin{eqnarray}
	U^{J\pi}_{fc,i} =-\frac{i}{\hbar} \langle \Psi^{JM\pi}_{f,c}(E_f) \vert V_{tr} \vert \Psi^{JM\pi}_i(E_i) \rangle,
\label{eq9}
\end{eqnarray}
where $V_{tr}$ is the transition potential. The scattering matrix can be reformulated as
\begin{align}
	U^{J\pi}_{if}=&-\frac{i}{\hbar}\int \chi^{J\pi}_{S_i L_i}(E_i,R) \,
	{\mathcal K}^{J\pi}_{L_i L_f}(R,R')\, \nonumber \\
&\times	\chi^{J\pi}_{S_f L_f,c}(E_f,R')dR dR',
	\label{eq9b}
\end{align}
where the transfer kernel is defined by
\begin{align}
		{\mathcal K}^{J\pi}_{L_i L_f}(R,R')=&{\mathcal J}\langle
	I^3(\pmb{r}'_x,\pmb{r}'_y)\, Y_{L_f}(\Omega_{R'})	 \nonumber \\
	&	\vert V_{tr} \vert
I_c^A(\pmb{r}_x,\pmb{r}_y)\, Y_{L_i}(\Omega_R)\rangle.
	\label{eq9c}
\end{align}
In this equation, ${\mathcal J}$ is the Jacobian, and $I^3$ and $I_c^A$ are the overlap integrals of the triton and of the residual nucleus ($^6$He or $^{11}$Li), respectively. We have omitted angular-momentum couplings for the sake of clarity. In Eq.\ \eqref{eq9c}, $\pmb{r_x}$, $\pmb{r_y}$, $\pmb{r'_x}$, and $\pmb{r'_y}$ are expressed as functions of $\pmb{R}$, and $\pmb{R'}$, and the integration is performed over $\Omega_R$ and $\Omega_{R'}$.

We closely follow the formalism presented in Ref.\ \cite{De21}, with two main differences: (1) the overlap integrals are obtained in a microscopic model; (2) core excitations are included, and provide the associated scattering matrices. From the scattering matrices, the transfer cross sections $d\sigma_t /d\Omega$ can be easily determined.

In the next sections, the RGM description of the $^6$He and $^{11}$Li nuclei is given in Ref.\ \cite{De23}. We adopt the same conditions of the calculations.

\section{The $\hept$ reaction}
Let us first consider the $\hept$ reaction which has been studied experimentally at $\elab=150$ MeV  \cite{WFR99,GRD05}. The $^6$He nucleus is a good candidate for a microscopic cluster study since the $\alpha$ core is strongly bound. It has been investigated in several works (see, for example, Refs.\ \cite{VSO94,FKK96,ASL99,BMB02} and references therein). 

\begin{figure}[htb]
	\centering
	\includegraphics[scale=0.45]{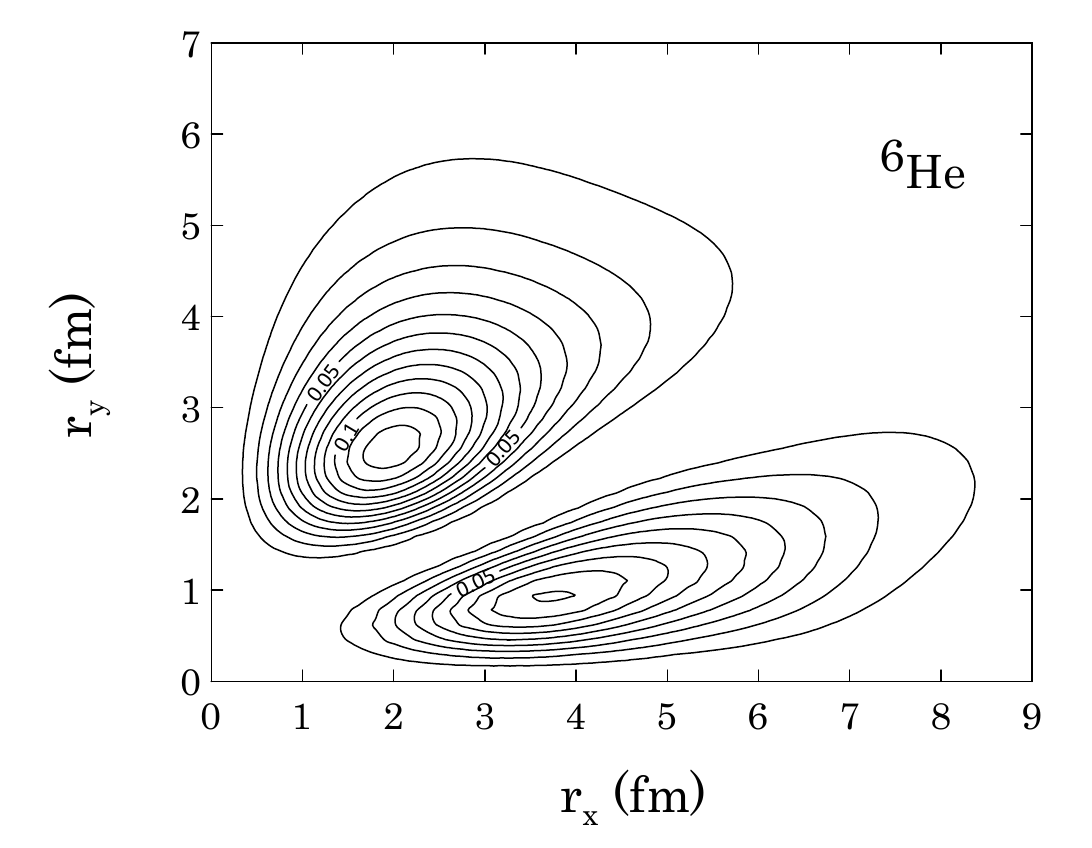}
	\caption{Radial amplitudes \eqref{eq5} of the $^{6}$He nucleus ($S_{12}=0$). The contour plots are drawn by steps of 0.01.}
	\label{fig_he6}
\end{figure}

In Fig.\ \ref{fig_he6}, we present the radial amplitude \eqref{eq5} for the neutron-neutron spin $S_{12}=0$. Given the spin-orbit force, a small  $S_{12}=1$ component is present (16.1\%). As it is well known, the radial amplitude presents two peaks associated with the so-called "dineutron" and "cigar" configurations \cite{BN10}. The spectroscopic factor $\mathcal{S}_c=1.41$ (see Table \ref{table1} and Eq.\ \eqref{eq6}) is in good agreement with the literature (1.39 \cite{VSO94}, 1.3957 \cite{BN10}, 25/16 \cite{Ti01}).

\begin{table}[h]
\caption{Spectroscopic factors of $^6$He and of $^{11}$Li for the core state $I_c$. }
\label{table1}
	\begin{tabular}{l|l}
	nucleus	& $\mathcal{S}_c$\\
		\hline
		$^6$He & \\
		$I_c=0^+$ & $   1.41$\\
		& \\
		$^{11}$Li & \\
		$I_c=3/2^-$ & $0.78\ (S_{12}=0), 0.13\ (S_{12}=1)$\\
		$I_c=1/2^-$ & $0.054\ (S_{12}=0), 0.050\ (S_{12}=1)$\\
	\end{tabular}
\end{table}

\begin{figure}[htb]
	\centering
	\includegraphics[scale=0.6]{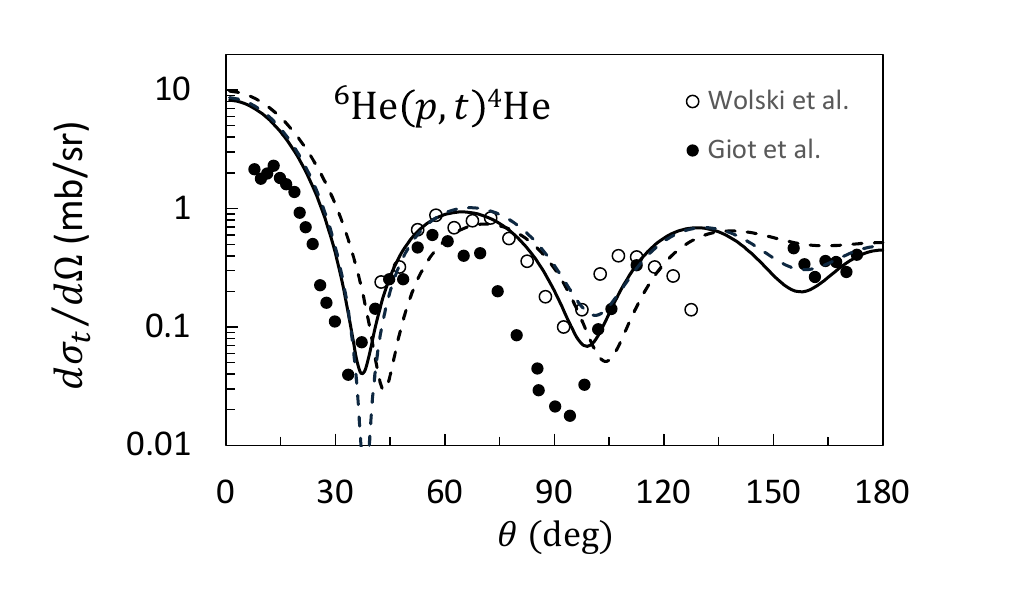}
	\caption{$\hept$ cross section with the GCM $^6$He overlap integral. The solid curve is obtained with the KD03 potential, and the dashed curves with the potentials given in Ref.\ \cite{WFR99}.}
	\label{fig_he6pt}
\end{figure}

The $\hept$ cross section $d\sigma/d{\Omega}_t$ is presented in Fig.\ \ref{fig_he6pt} with the data of Wolski {\sl et al.}\ \cite{WFR99} and of  Giot {\sl et al.} \cite{GRD05}. Both data sets are obtained with an $^6$He radioactive beam at $\elab = 150$ MeV. To test the sensitivity to the $\hep$ optical potential, we have used three variants: the compilation of Koning and Delaroche \cite{KD03} (referred to as KD03), and the two potentials proposed in Ref.\ \cite{WFR99} which have been fitted to elastic-scattering data. In the exit channel, the $\alpha+t$ potential of Giot et al.\ \cite{GRD05} (pot $B$) is used.

Figure \ref{fig_he6pt} shows that there is a slight sensitivity to the $\hep$ potential, but the general trend is similar. The sensitivity to the $\alpha+t$ potential is still weaker. The RGM overestimates the data at small angles, but the general agreement is fair. In particular the locations of the minima are well reproduced. The calculation is in better agreement with the data of Wolski {\sl et al.} \cite{WFR99} than with those of Giot {\sl et al.} \cite{GRD05}.

The calculation of the transfer cross sections involves several inputs (optical potentials, overlap integrals) and cannot be expected to provide an excellent agreement with experiment. In addition, the present microscopic method for the overlap integrals avoids the introduction of a spectroscopic factor, usually used in DWBA calculation to fit experimental data.

\section{The $\lipt$ reaction}
\begin{figure}[htb]
	\centering
	\includegraphics[scale=0.45]{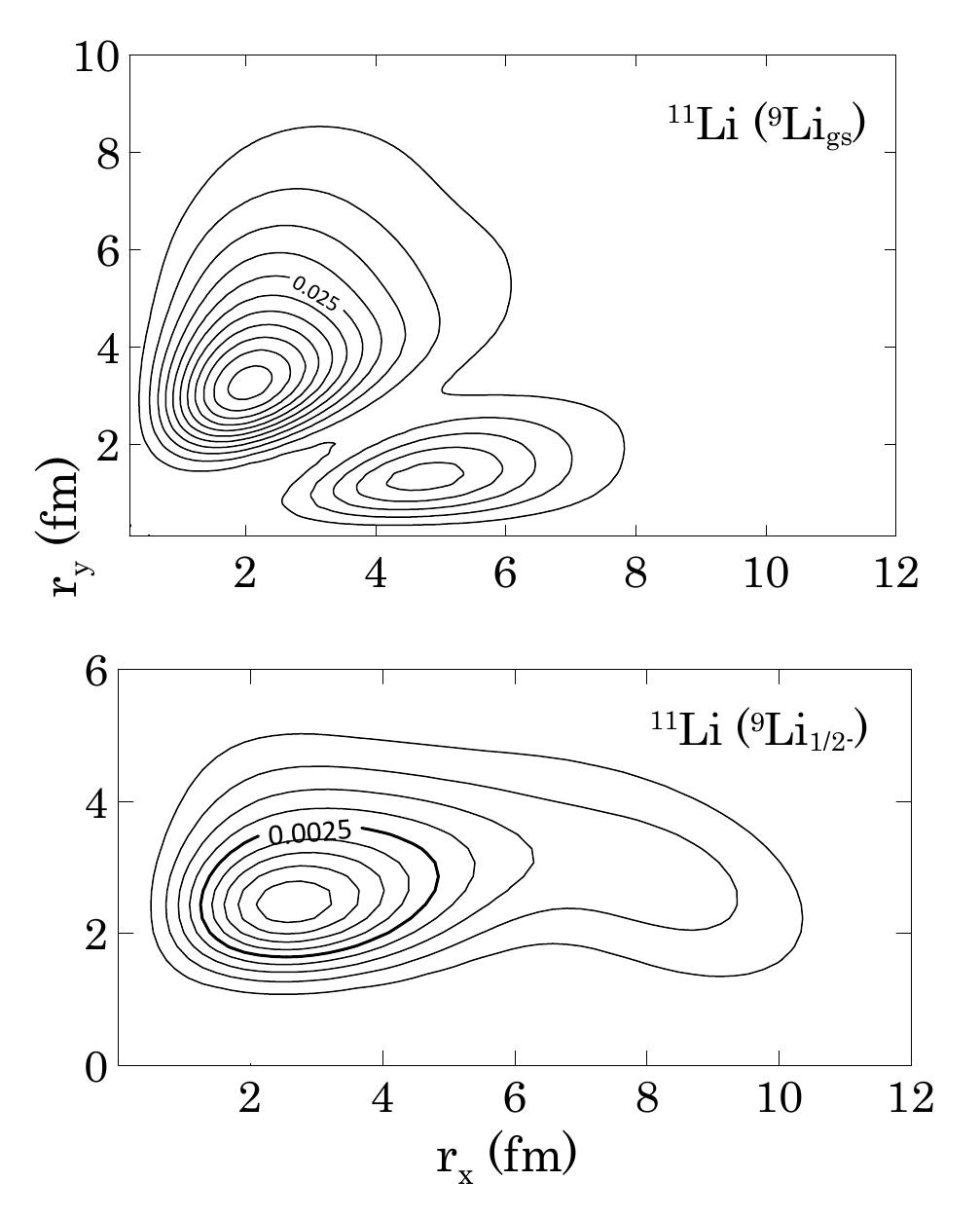}
	\caption{Radial amplitudes \eqref{eq5} for the $^{11}$Li nucleus for $S_{12}=0$ and for the $^9$Li ground state (top) and $1/2^-$ excited state (bottom). The contour plots are drawn by steps of 0.005 and 0.0005.}
	\label{fig_li11}
\end{figure}

The $^{11}$Li radial amplitudes are presented in Fig.\ \ref{fig_li11}, where we illustrate the ground-state component and the $^9$Li($1/2^-$) configuration. The main contribution $S_{12}=0$ is shown. Table \ref{table1} presents the spectroscopic factors in various channels. As expected, the main component is associated with the $^9$Li ground state (0.78 and 0.13 for $S_{12}=0$ and $S_{12}=1$, respectively). The spectroscopic factors in the excited configuration are much smaller (0.054 and 0.050).

\begin{figure}[htb]
	\centering
	\includegraphics[scale=0.6]{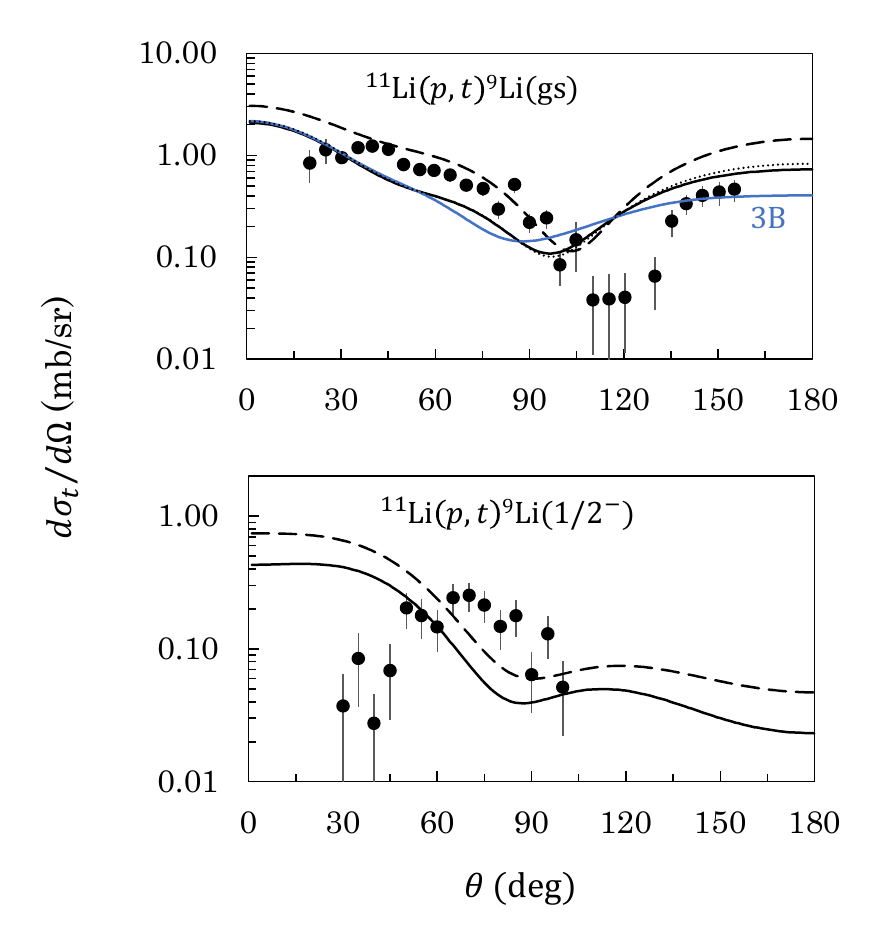}
	\caption{$\lipt$ cross sections. The solid lines are obtained with the $\lip$ potential of Ref.\ \cite{De21} and the dashed lines with the KD03 global potential. The dotted line corresponds to the $^{11}$Li overlap integral without core excitation. The three-body calculation of Ref.\ \cite{De21} is shown in blue and is labeled as `3B'. The data are taken from Ref.\ \cite{TAB08}.}
	\label{fig_lipt}
\end{figure}

The $\lipt$ cross sections are shown in Fig.\ \ref{fig_lipt}  with the data of Tanihata {\sl et al.} \cite{TAB08}. The present microscopic theory not only provides the ground-state cross section, but also the contribution of the $^9$Li($1/2^-$) first excited state. This cross section has been measured in Ref.\ \cite{TAB08} and is expected to probe the $^{11}$Li structure. As there are no elastic-scattering data at the corresponding energy ($\elab = 33$ MeV), we use two optical potentials. An equivalent CDCC potential \cite{De20}, which is obtained with a four-body description of the system, is used. To assess the sensitivity of the cross section to this input, we also use the global parametrization KD03. For the exit $\lit$ channel, we adopt the potential of Pang {\sl et al.} \cite{PRS09}.

For the ground-state contribution, the cross section at small angles is in fair agreement with experiment, which shows that the overlap integral obtained in the RGM is reliable. The present model reproduces a minimum near $\theta=120^{\circ}$, but overestimates the data at large angles. We present as a dotted line the cross section with a single-channel description of $^{11}$Li, which shows that core excitations do not play an important role in this reaction. For the sake of comparison, the cross section obtained in Ref.\ \cite{De21} with a three-body description is also shown. This non-microscopic approach involves a $^9$Li+n potential which is poorly known, in contrast with the present model where the $^{11}$Li wave functions and overlap integrals are derived from a nucleon-nucleon interaction.

The potential of Ref.\ \cite{De20} is obtained from a CDCC calculation involving the $^{11}$Li breakup, and is expected to be more reliable than the global parametrization of KD03 (dashed line in Fig.\ \ref{fig_lipt}) which is fitted at higher energy and on heavier nuclei. It is, however, widely used since many nucleon-nucleus optical potentials are not accurately known.  A measurement of the elastic cross section at the same energy would be welcome in the determination of the transfer cross section.

The $\lipt^*$ cross section is smaller than the ground state contribution, as expected from experiment and from the different overlap integrals (see Fig.\ \ref{fig_li11}). In the angular range $60^{\circ} < \theta <120^{\circ}$, the calculation is consistent with experiment. At small angles, however, the behavior is different. The data present a fast decrease for $\theta < 60^{\circ}$ whereas the semi-microscopic cross section weakly depends on the angle. 

\section{Energy dependence of the cross sections}
\begin{figure}[htb]
	\centering
	\includegraphics[scale=0.6]{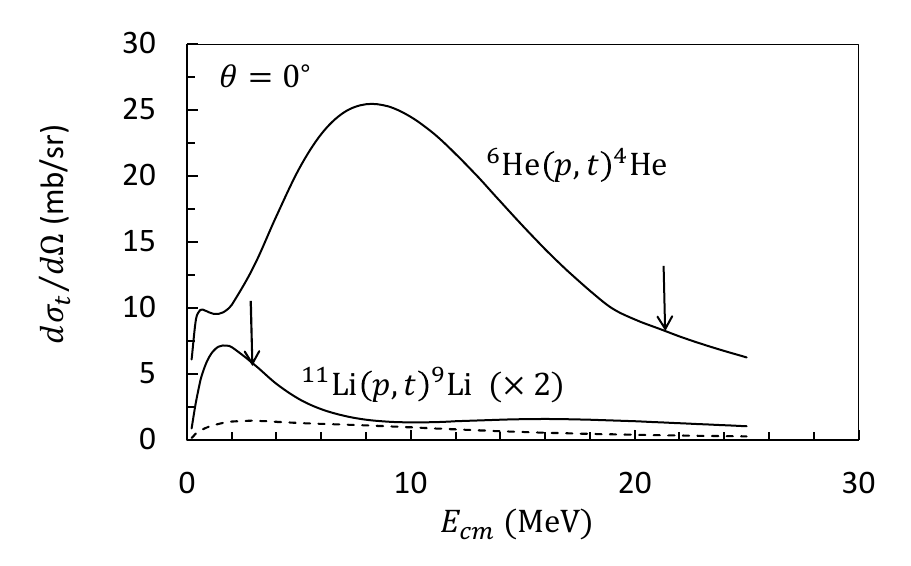}
	\caption{$\hept$ and $\lipt$ cross sections at $\theta=0^{\circ}$ as a function of energy. For $\lipt$ the dashed line corresponds to the $^9$Li excited state.}
	\label{fig_zero}
\end{figure}

For both considered reactions, data are available at a single energy. To investigate the energy dependence we show in Fig.\ \ref{fig_zero} the cross section at $\theta=0^{\circ}$ in the energy interval $\ecm < 25$ MeV. We indicate by arrows the experimental energies. As found in the Ref.\ \cite{De21} in a non microscopic description of $^{11}$Li, the energy adopted in Ref.\ \cite{TAB08} nearly corresponds to the maximum cross section. For the $\hept$ reaction, however, the cross section is predicted to increase at lower energy. The maximum is found near $\ecm \approx 8$ MeV, i.e. at $\elab \approx 56$ MeV. In both reactions the data availability at other energies would be a valuable test of the model. Two neutron transfer cross sections on other halo nuclear such as $^{14}$Be would also be welcome. 

\section{Conclusion and outlook}
This paper explores the structure of halo nuclei, such as $^{6}$He and $^{11}$Li, through $(p,t)$ reactions, which offer insights into their spectroscopic properties.  We employ a semi-microscopic model using the Resonating Group Method (RGM) to calculate overlap integrals, incorporating the Pauli principle and allowing for core excitations. This approach does not contain any fitting parameter, and is particularly suitable for halo nuclei due to their core + neutron + neutron structure.

The RGM generates microscopic overlap integrals used as input for calculating $(p,t)$ cross sections within the Distorted Wave Born Approximation (DWBA).  The model is applied to both $\hept$ and $\lipt$ reactions.  For $^{6}$He, the calculated cross section reasonably matches experimental data, although it overestimates values at small angles.  The results demonstrate a low sensitivity to the choice of the $\hep$ optical potential.

In the case of  $^{11}$Li, the model predicts both the ground state and excited state contributions to the cross section. The spectroscopic factors for $^{11}$Li are calculated for various channels, revealing the dominant role of the  $^{9}$Li ground state configuration.  The $\lipt$ cross section calculations are compared to experimental data, showing reasonable agreement.  The inclusion of core excitations and the use of microscopic overlap integrals within the RGM framework offer a valuable tool for investigating the structure of halo nuclei through $(p,t)$ reactions.  

The dependence of the cross section at $\theta=0^{\circ}$ has been analyzed, and suggests that the $\hept$ cross section could be larger and at lower energy ($\elab \approx 50$ MeV). Generally speaking, measurements at other energies would be useful to assess theoretical models.

The present model goes beyond traditional DWBA calculations, where the projectile and residual-nucleus wave functions are described in the simple potential model, neglecting the internal structure. Of course a fully microscopic approach would be desirable, but is currently not feasible, essentially for two reasons: $(i)$ we are considering weakly bound halo nuclei and a scattering model should include these properties. A microscopic cluster study of $^6$He and of $^{11}$Li is possible \cite{De19}, but is very demanding in terms of computer capabilities. Adding an incident particle means that a double angular-momentum projection is necessary which would still considerably increase the computer times. $(ii)$ More important, a reaction such as $\lip$, even at low energies, presents many open channels  which cannot be accounted for in a microscopic theory. The use of optical potentials, which simulate the absorption to these open channels, is therefore unavoidable.

\section*{Acknowledgments}
This work was supported by the Fonds de la Recherche Scientifique - FNRS under Grant Numbers 4.45.10.08 and J.0065.22.


\end{document}